\documentclass[a4paper]{jpconf}
\usepackage{graphicx}
\usepackage{cite}

\begin{document}
\title{Moments Method for Shell-Model Level Density}

\author{V Zelevinsky$^1$, M Horoi$^2$, and R A Sen'kov$^2$}

\address{$^1$ Department of Physics and Astronomy and National Superconducting Cyclotron Laboratory, Michigan State University, East Lansing, MI 48824-1321,USA}

\address{$^2$ Department of Physics, Central Michigan University, Mount Pleasant, MI 48859, USA}

\ead{zelevins@nscl.msu.edu}

\begin{abstract}
The modern form of the Moments Method applied to the calculation of the nuclear shell-model
level density is explained and examples of the method at work are given. The calculated level
density practically exactly coincides with the result of full diagonalization when the latter
is feasible. The method provides the pure level density for given spin and parity with spurious
center-of-mass excitations subtracted. The presence and interplay of all correlations leads to
the results different from those obtained by the mean-field combinatorics.
\end{abstract}

The long-standing problem of nuclear level density is important both from a theoretical viewpoint
as a powerful instrument for studying nuclear structure and for numerous applications. Nuclear
technology requires the detailed knowledge of level density, especially in the region around the
neutron separation energy. Astrophysical reactions responsible for the nucleosynthesis in the
Universe can be understood only if we know the nuclear level density in the Gamow window. The
interest to this problem is also connected with the necessity to develop methods of statistical
physics and thermodynamics for small {\sl mesoscopic} systems (complex nuclei, atoms and molecules,
nanoscale devices, prototypes of quantum computers).

Application of statistical physics to studies of nuclear properties was started in the early years
of nuclear physics by Frenkel \cite{frenkel36}, Bethe \cite{bethe36,bethe37} and Landau \cite{landau37}.
Such ideas served as a base for a concept of compound nucleus by Bohr \cite{bohr36}. The review of
those ideas and their further development was given in Refs. \cite{ericson60,gilbert65}. At this stage
theory of level density and detailed microscopic nuclear theory were growing more or less independently
along parallel paths. Their synthesis became possible when the reliable versions of the nuclear
shell model were established for several segments of the nuclear chart. Of course, this is related
to the modern computational tools.

In this presentation we show the practical way of finding the density of energy levels with fixed
exact quantum numbers (spin, parity, isospin projection) for a given shell model Hamiltonian. The method
becomes practical because it is avoiding the full diagonalization of prohibitively large matrices.
In contrast to popular mean-field approaches \cite{goriely08,goriely09,nld2}, all many-body correlations
are fully accounted for, albeit in a restricted orbital space. The exact classification of states according
to the conserved quantities and therefore the absence of artificial standard spin cut-off parameters
is the advantage compared to Monte-Carlo calculations although some of them give relatively good results
\cite{alhassid07,alhassid08}. Similar problems still may arise in approaches based on statistical spectroscopy
\cite{teran06}.

Our method was gradually developed in several publications \cite{HJZ02,HKZ03,HGZ04,HZ07,gao09,gao09a,scott10,
SH10,SHZ11}. A high-performance code for calculating spin- and parity-dependent level density is available via
the Computer Physics Communication homepage on Science Direct \cite{SHZcomp13}. Technically we use
the {\sl Moments method} based on statistical spectroscopy \cite{wong86,kota10}, the applications are compared
to the results obtained with the full diagonalization in the nuclear shell model with effective interactions
when such calculations are practically possible \cite{big}. The method is reinforced by the use of the recurrence
relations \cite{jacquemin81,isacker02,HZ07} for eliminating spurious contributions.

We start with the shell-model Hamiltonian that contains mean-field single-particle energies in the restricted
space and effective two-body interactions (many-body forces can be added without problems although the equations
become more cumbersome). The spherical orbitals $|njm\rangle$ are used as a basis, and the counted many-body levels 
have definite total spin $J$ and parity. A certain distribution of non-interacting particles over orbitals
defines a configuration ({\sl partition}). For each partition $p$, let $D_{\alpha p}$ be the number of many-body
states with exact quantum numbers $\{\alpha\}$ including the number of particles, spin, parity, isospin...
The {\sl diagonal} matrix elements of the total (including interaction) Hamiltonian define the energy {\sl centroid}
$E_{\alpha p}$ for the states of class $\alpha$ built in the partition $p$,
\begin{equation}
E_{\alpha p}=\,\frac{1}{D_{\alpha p}}\,{\rm Tr}^{(\alpha p)}\,H.                            \label{1}
\end{equation}
For each partition we define also the effective {\sl width} (energy dispersion) $\sigma_{\alpha p}$,
\begin{equation}
\sigma_{\alpha p}^{2}=\,\frac{1}{D_{\alpha p}}\,{\rm Tr}^{(\alpha p)}\,H^{2}-(E_{\alpha p})^{2}. \label{2}
\end{equation}

The contribution of each state to this width, related to the strength function of an unperturbed state fragmented
over the exact eigenstates, can be found summing the squares of off-diagonal matrix elements of the full Hamiltonian
over a given row of the matrix $H$. Here the interaction between the states of different partitions is fully
accounted for. As known from the shell-model experience \cite{big,pillet12}, the centroids of partitions form
a smooth sequence, while the widths of individual unperturbed states fluctuate only weakly which can be considered
as one of features of quantum chaos (in particular, {\sl geometric chaoticity} of angular momentum addition
\cite{ZVRep}, that emerges, for a realistic strength of interactions, already at rather low excitation energy).
The details of calculating the trace of $H^{2}$ and explanation of the computational algorithm can be found in Ref.
\cite{SHZcomp13}; a computational method based on group theory but close in spirit to our approach is presented in Ref. \cite{launey13}.

The set of quantities (\ref{1}) and (\ref{2}) forms a foundation of the Moments Method \cite{wong86}.
The level density of each partition is known \cite{brody} to be close to Gaussian. The first two moments
define it with sufficient precision. The total level density is restored by summing the contributions
of the partitions,
\begin{equation}
\rho(E;\alpha)=\sum_{p}D_{\alpha p}G_{\alpha p}(E),                                     \label{3}
\end{equation}
where $G_{\alpha p}(E)$ is a {\sl finite range Gaussian} with the centroid at $E_{\alpha p}$ counted
from the ground state energy; this Gaussian is cut at a distance $\eta\sigma_{\alpha p}$ from the centroid,
where $\eta$ is an empirical parameter eliminating the unphysical tails of the Gaussian (there are
theoretical arguments confirmed by experience that fix $\eta$ between 2.5 and 3). The ground state energy
is to be found independently, for example by the approximate diagonalization in a smaller space and
the exponential convergence method \cite{ECM99,ECM02}, where the exponential regime also starts
approximately at a distance of 3-4 spreading widths.

The last essential point is exclusion of unphysical states corresponding to the center-of-mass excitation
which are present in the shell-model calculations due to cross-shell mixing. If we classify the basis states
similarly to the harmonic oscillator field by $N\hbar\omega$ excitations, then for $N=0$ the spurious excitations
are absent, $\rho_{{\rm pure}}(E,J,N=0)=\rho_{{\rm calc}}(E,J,N=0)$. With $N=1$ excitations we have to subtract
from the shell-model result the states which could be obtained by the action of the center-of-mass vector operator
on all states of $N=0$ class,
\begin{equation}
\rho_{{\rm pure}}(E,J,N=1)=\rho_{{\rm calc}}(E,J,N=1)-\sum_{J'=J,J\pm 1}\rho_{{\rm calc}}(E,J',N=0). \label{4}
\end{equation}
The final result is expressed by a recurrent relation similar to used in \cite{jacquemin81,isacker02}:
\begin{equation}
\rho_{{\rm pure}}(E,J,N)=\rho_{{\rm calc}}(E,J,N)-\sum_{K=1}^{N}\sum_{J_{K}=J_{{\rm min}}}^{N,{\rm step}\,2}
\sum_{J'=|J-J_{K}|}^{J+J_{K}}\rho_{{\rm calc}}(E,J',N-K).                                       \label{5}
\end{equation}
\begin{figure}[h]
\begin{minipage}{0.45\linewidth}
\includegraphics[width=1.0\linewidth]{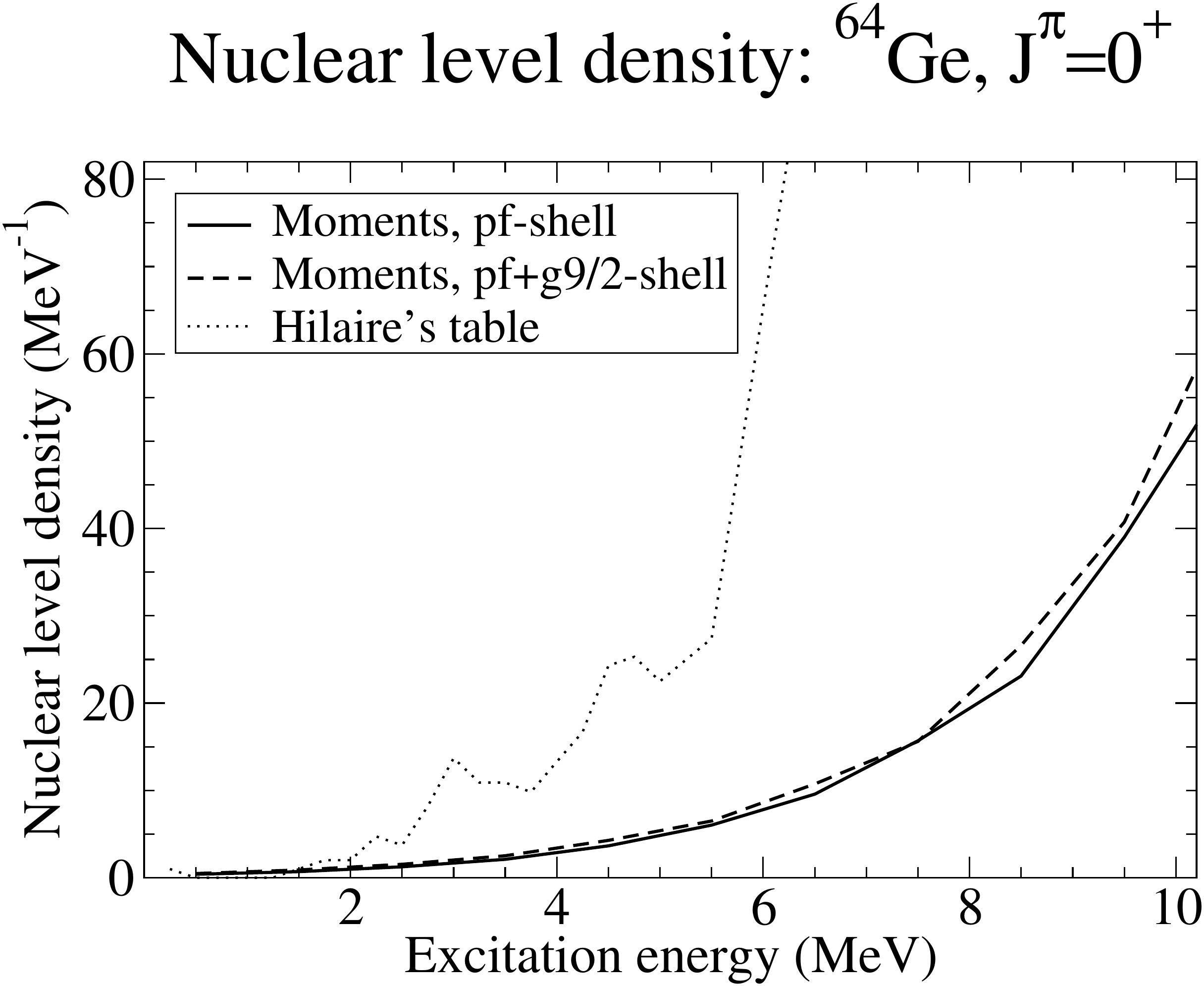}
\end{minipage}\hspace{2pc}%
\begin{minipage}{0.45\linewidth}
\includegraphics[width=1.0\linewidth]{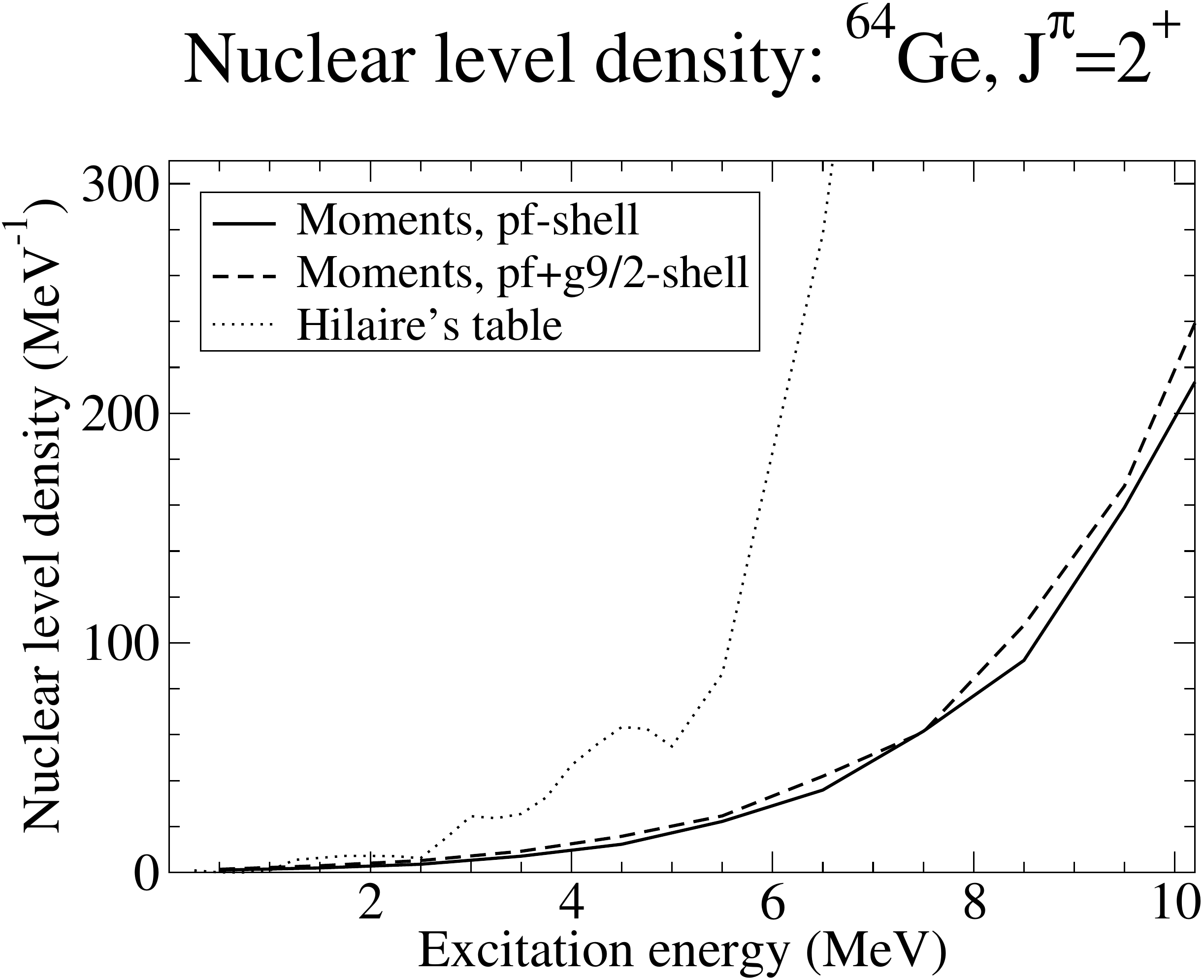}
\end{minipage}
\caption{\label{fig1} $^{64}$Ge nuclear level density for $J=0,2$ and positive parity. Solid and dashed lines present
calculations by Moments method in $pf$- (calculated with GXPF1A interaction) and $pf+g_{9/2}$-model spaces correspondingly.
Dotted lines present level densities of Ref. \cite{nld2}. }
\end{figure}

\begin{figure}[h]
\begin{minipage}{0.45\linewidth}
\includegraphics[width=1.0\linewidth]{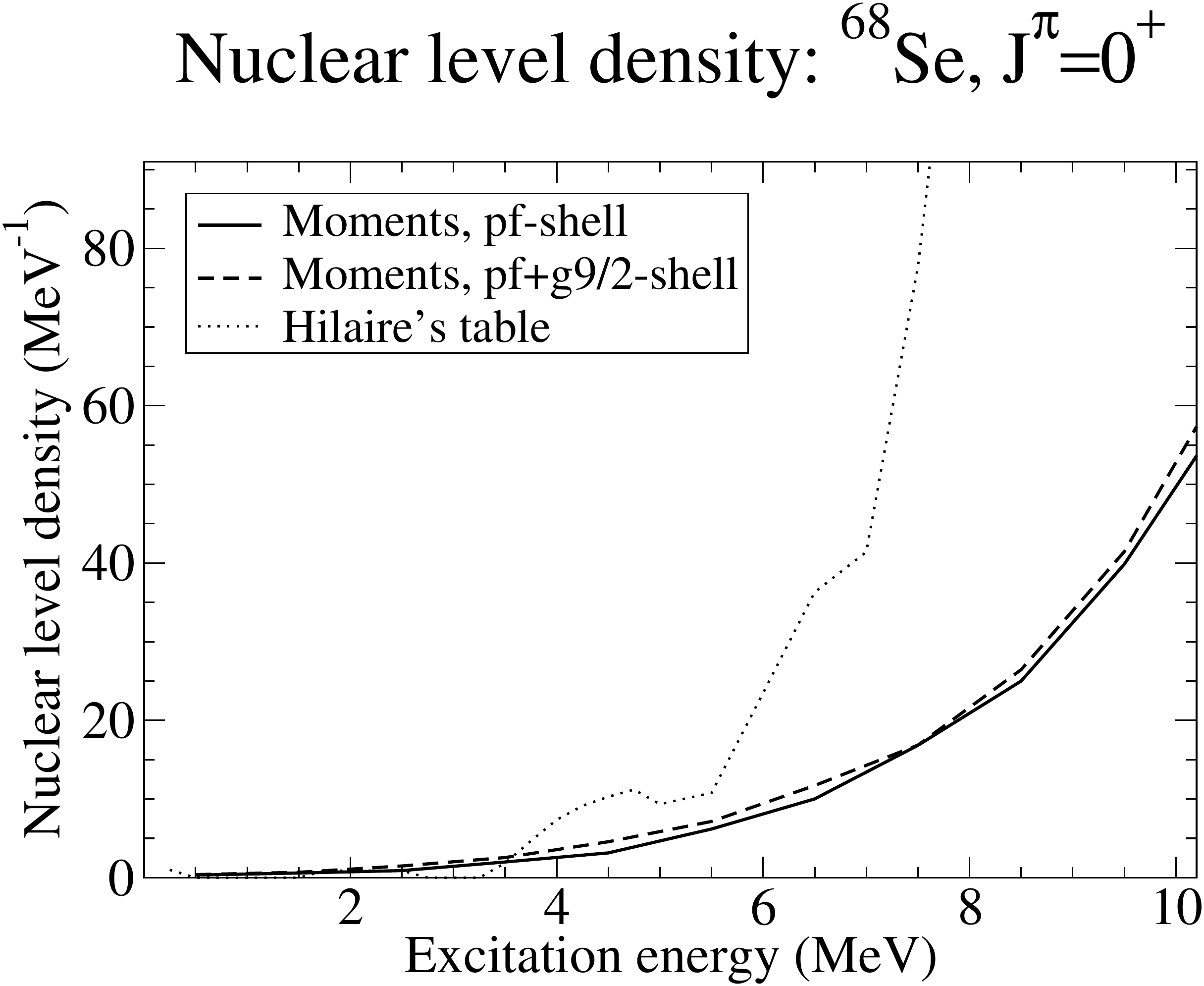}
\end{minipage}\hspace{2pc}%
\begin{minipage}{0.45\linewidth}
\includegraphics[width=1.0\linewidth]{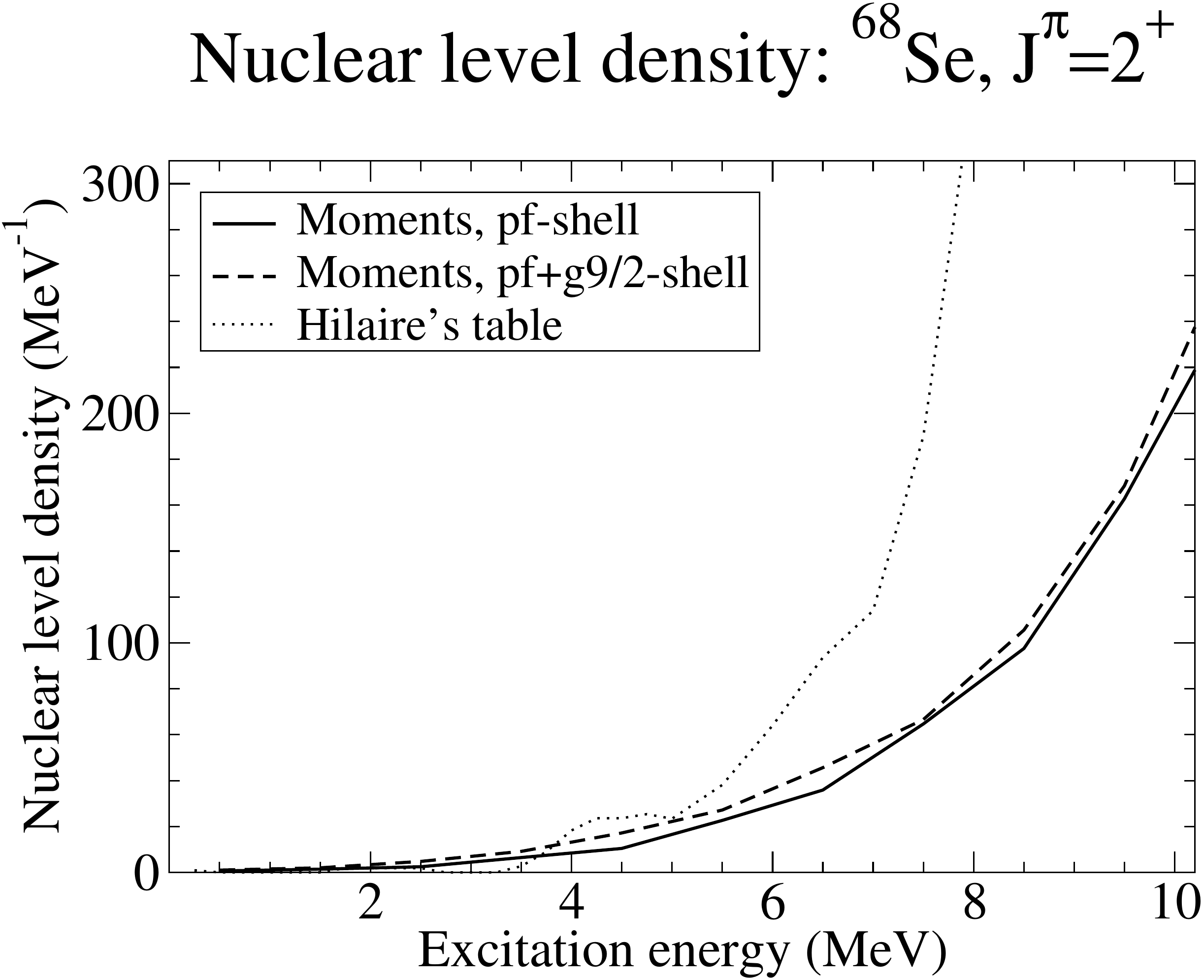}
\end{minipage}
\caption{\label{fig2} The same as Fig. 1 for $^{68}$Se.}
\end{figure}

We illustrate the results by applications of the method to several nuclei important for astrophysics.
Fig. 1 shows the level density for the $N=Z$ nucleus $^{64}$Ge, states $J^{\pi}=0^{+}\;{\rm and}\;2^{+}$.
Fig. 2 gives the same for $^{68}$Se. In all four cases, the results for a larger orbital space, $pf+g_{9/2}$
shells,
only slightly exceed those for the $pf$ space. However, as it is typical for all our calculations, the shell-model
level density is very smooth as a function of energy and turns out to be much lower than what can be found from the
tables \cite{nld2}. The preliminary conclusion is that the full set of effective shell-model interactions
responsible for emergence of quantum chaos is very important smoothing the results based mainly on the mean-field
combinatorics.

\begin{figure}[h]
\begin{minipage}{0.45\linewidth}
\includegraphics[width=1.0\linewidth]{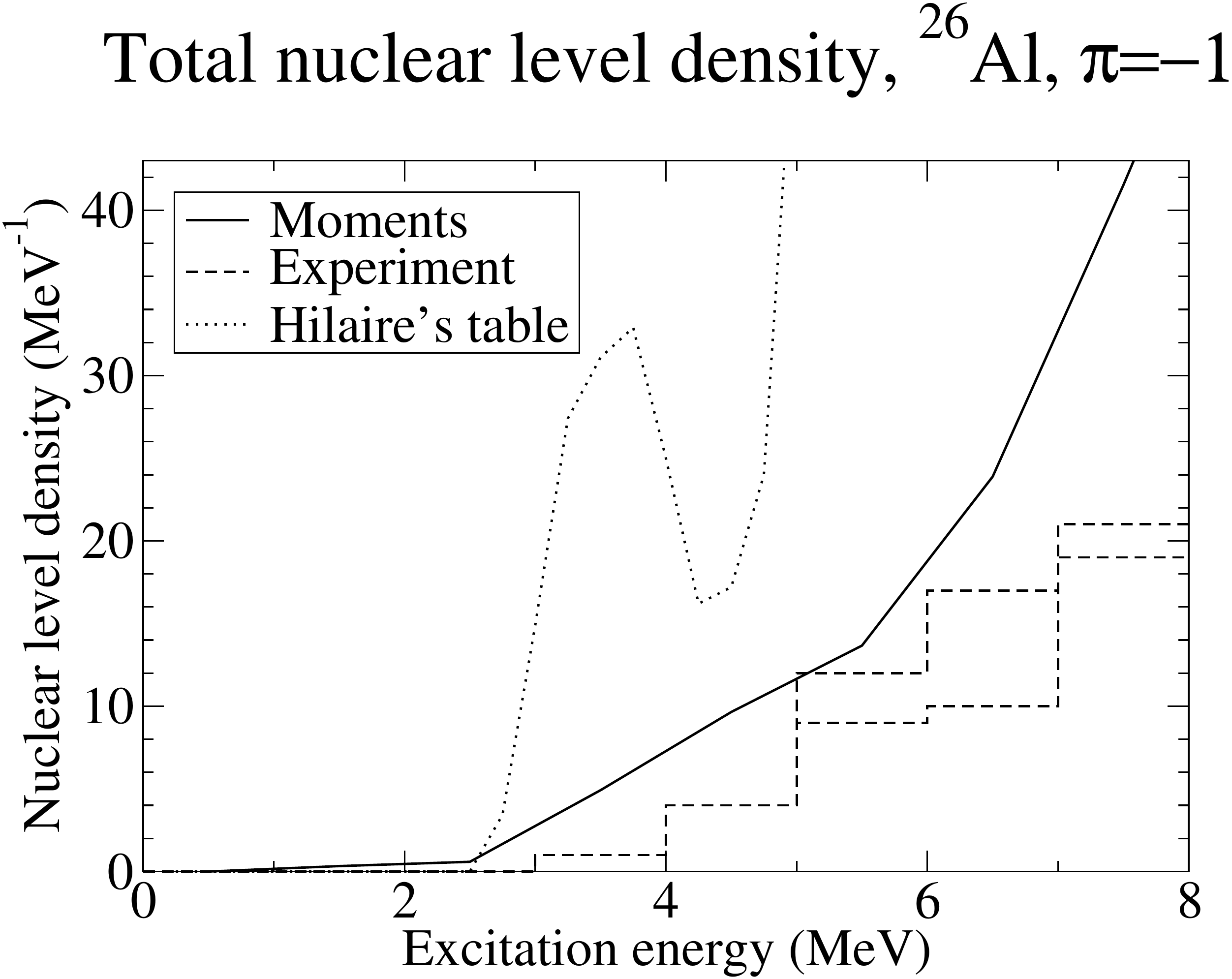}
\end{minipage}\hspace{2pc}%
\begin{minipage}{0.45\linewidth}
\includegraphics[width=1.0\linewidth]{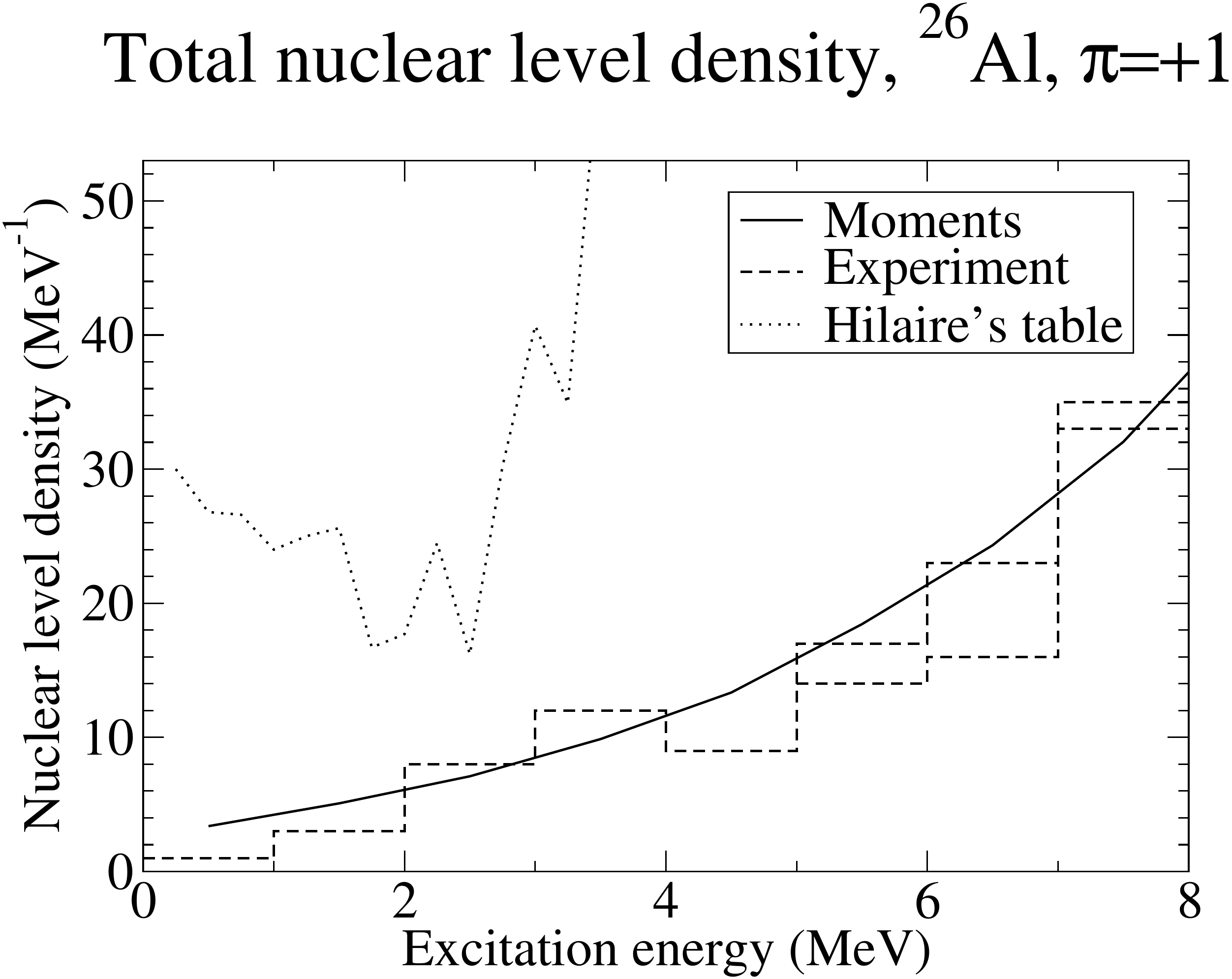}
\end{minipage}
\caption{\label{fig3} $^{26}$Al, {\it s-p-sd-fp}-model space, all spins, WBT interaction \cite{wbt}. Nuclear level densities are presented for negative (left figure) and positive (right figure) parities. Solid lines present the results of the Moments method, dashed lines show experimental level density (upper and lower lines correspond to experimental uncertainties in level identification), and dotted lines present the total level densities of Ref. \cite{nld2}. }
\end{figure}

In Fig. 3 we compare the total level density (for positive and negative parity separately) in $^{26}$Al, calculated in
the large {\it s-p-sd-fp}-model space restricted by $1\hbar\omega$ excitations only, with the WBT interaction \cite{wbt}.
For this isotope a good set of data is available and it is possible to juxtapose the results of our approach to those
of the full shell-model diagonalization. Again, we obtain much lower and smoother level density than predicted
by tables \cite{nld2}. The two dashed stair-case lines show the data (``optimistic" and ``pessimistic"
estimates of uncertainties in identification of level parity). The calculations reasonably well describe
available data.

\begin{figure}[h]
\begin{minipage}{0.45\linewidth}
\includegraphics[width=1.0\linewidth]{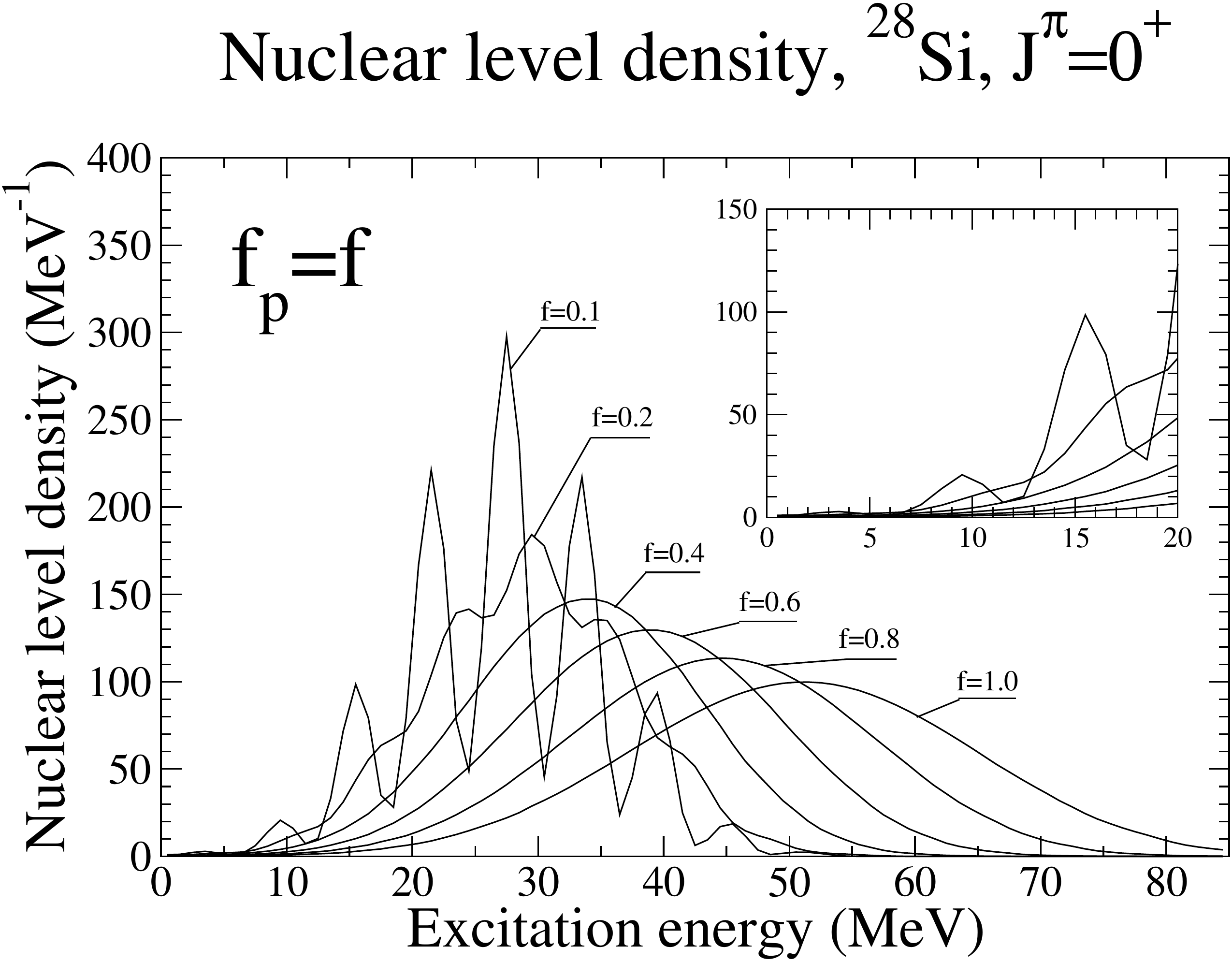}
\end{minipage}\hspace{2pc}%
\begin{minipage}{0.45\linewidth}
\includegraphics[width=1.0\linewidth]{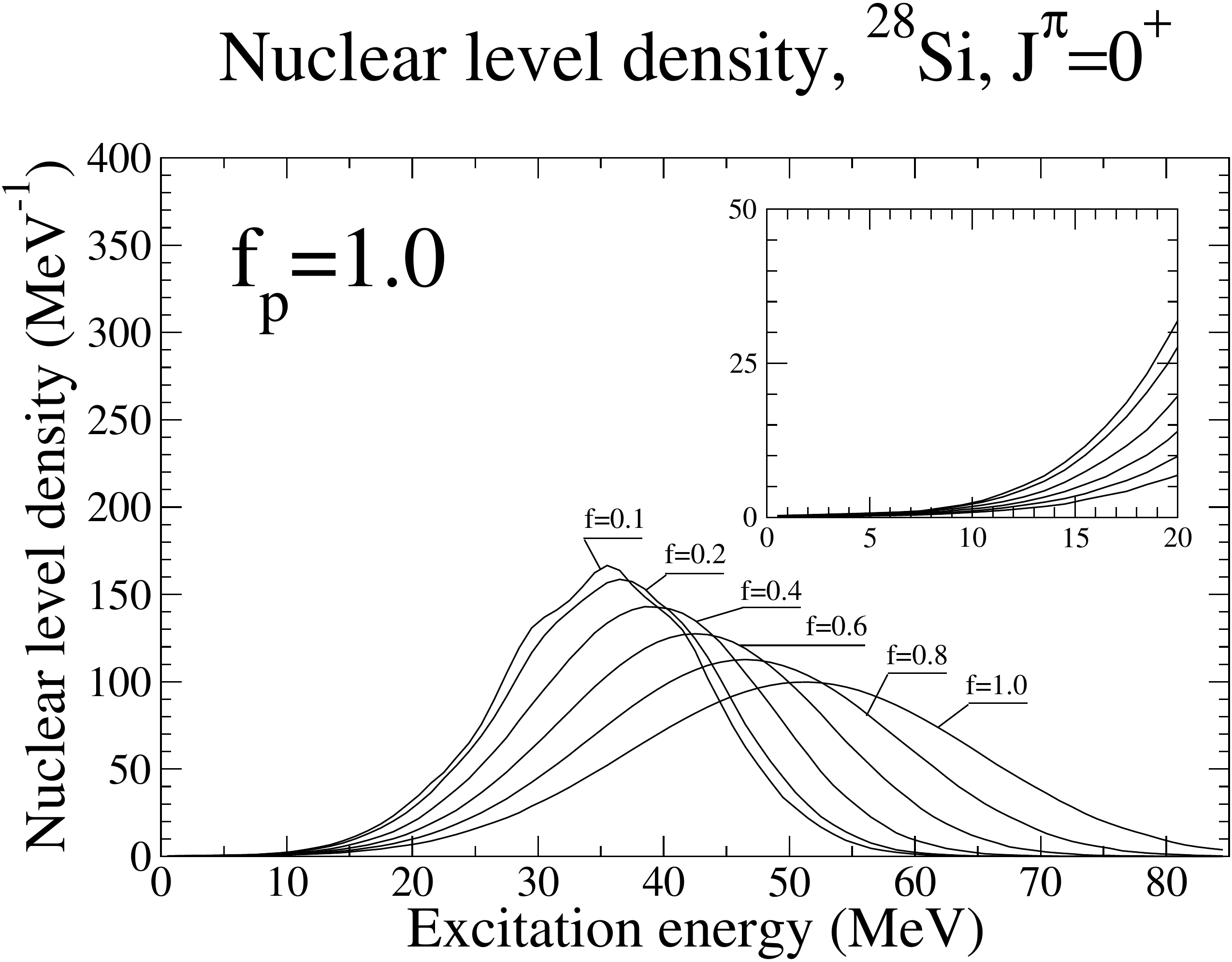}
\end{minipage}
\caption{\label{fig4} $^{28}$Si, $J^\pi=0^+$, $sd$-model space,
USDB interaction \cite{usd}.
Different curves correspond to different scale factors $f$,
see Eq. (\ref{6}). Left panel presents the case when $f_p=f$, i.e.
pairing and non-pairing parts of the interaction scales
similarly. Right figure presents the case when pairing interaction
is always on, $f_p=1$, while non-pairing interaction scales. The inset shows the low-energy part.
}
\end{figure}

\begin{figure}[h]
\begin{minipage}{0.45\linewidth}
\includegraphics[width=1.0\linewidth]{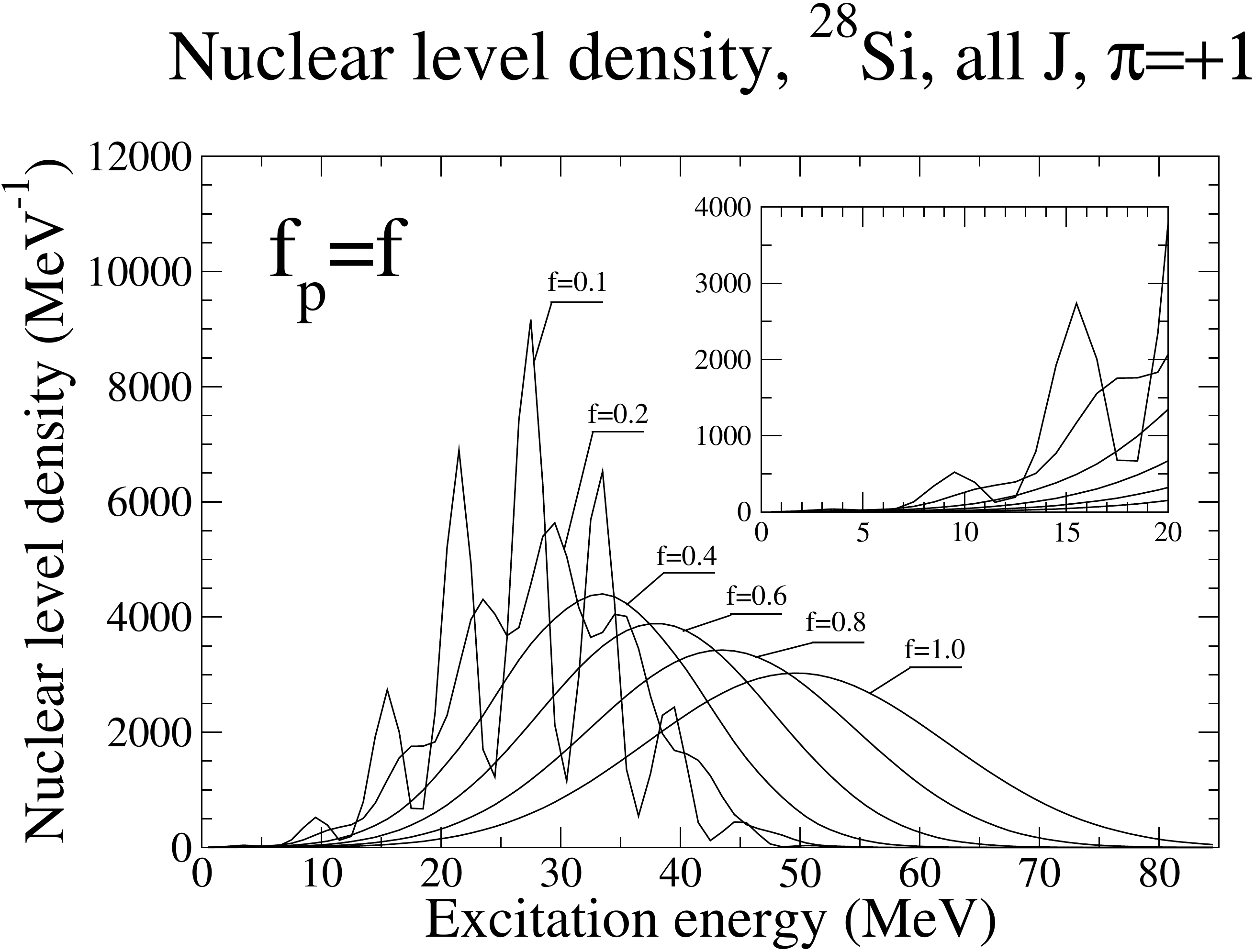}
\end{minipage}\hspace{2pc}%
\begin{minipage}{0.45\linewidth}
\includegraphics[width=1.0\linewidth]{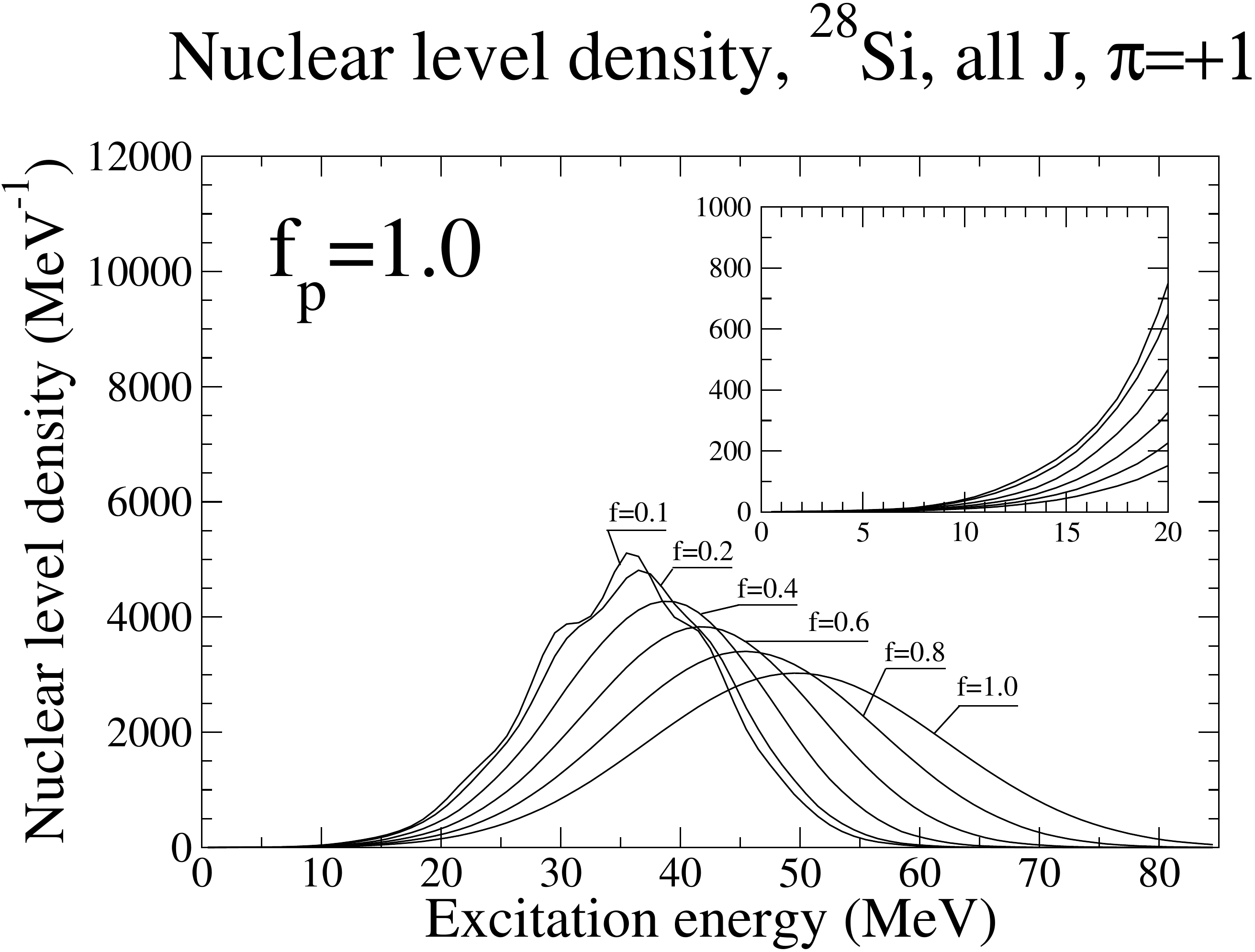}
\end{minipage}
\caption{\label{fig5} $^{28}$Si, all $J$, $\pi=+1$, $sd$-model space. The description is similar to Fig. \ref{fig4}.}
\end{figure}

Using the Moments method we can study dependence of the
level densities on the interaction. Let us scale
the shell model two-body interaction with two factors,
$f$ and $f_p$:
\begin{equation} \label{6}
H = \epsilon + f_p V(\mbox{pairing}) + f V(\mbox{non-pairing}),
\end{equation}
where $\epsilon$ is the one-body part of the Hamiltonian, $V(\mbox{pairing})$ includes only
the $J=0, T=1$ matrix elements of the two-body interaction, while $V(\mbox{non-pairing})$
includes all remaining matrix elements.
Varying the scale factors $f$ and $f_p$ we can study the influence of different components
of interaction onto the level density. Figs. \ref{fig4} and \ref{fig5}
present examples of such an investigation for $^{28}$Si
level densities, calculated in the $sd$-model space,
with the USDB interaction \cite{usd}.
Fig. \ref{fig4} presents results only for $J=0$ and Fig. \ref{fig5}
presents analogous calculations for all spins. Although dimensions and some details depend on $J$,
the qualitative behavior is universal in all classes of states. The sharp irregularities related to the partition
structure are rapidly smoothen even by the pairing interaction when it is accounted for exactly rather than
in the BCS or HFB schemes. The remaining interactions broaden the level density and made its final tuning.

The natural next step in development of theory should be an attempt, in application to many nuclei, to extract
the global empirical parameters, such as temperature, parameters of phenomenological fits, spin cut-off, and
so on in order to bridge the gap between shell-model theory and convenient statistical descriptions used
by experimentalists. Such studies also could help in selection of more reliable effective interactions for
different regions of the periodic table. It might be that the global behavior of the level density is governed
by few ``coherent" interactions while the remaining abundant matrix elements are responsible for broadening and smoothing
the stair-case behavior and can be modeled by an appropriately chosen ``noise". With parallelization of calculations \cite{SHZcomp13}, they can be extended to larger many-shell spaces. Finally, we have to confess that
the problem of continuum effects is still waiting for its proper consideration.

\section*{Acknowledgments}
V.Z. and M.H. acknowledge support from  the NSF grant PHY-1068217, the work of R.A.S and M.H.
was supported by the DOE NUCLEI grant DE-SC0008529. V.Z. is grateful to S. Goriely and H. Schatz
for constructive discussions during the NPA6 conference.

\end{document}